\documentclass[showpacs,preprintnumbers]{revtex4}
\usepackage{amssymb}
\usepackage{amsmath}
\usepackage{dcolumn}
\usepackage{graphicx}

\begin{document}

\title{Effect of bulk viscosity on Elliptic Flow near QCD phase transition}
\author{G. S. Denicol${}^{a}$, T. Kodama${}^{a}$, T. Koide${}^{b}$, and Ph.
Mota${}^{a}$}
\affiliation{$^{a}$Instituto de F\'{\i}sica, Universidade Federal do Rio de Janeiro, C.
P. 68528, 21945-970, Rio de Janeiro, Brasil}
\affiliation{$^{b}$FIAS, Johann Wolfgang Goethe-Universit\"at, Ruth-Moufang Str. 1,
60438, Frankfurt am Main, German}

\begin{abstract}
In this work, we examine the effect of bulk viscosity on elliptic flow taking
into account the critical behavior of the EoS and transport coefficients
near the QCD phase transition. We found that the $p_{T}$ dependence of $v_{2}
$ is quantitatively changed by the presence of the QCD phase transition.
Within reasonable values of the transport coefficients, $v_{2}$ decreases by
a factor of $15\%$ at low $p_{T}$ ($<1$ GeV). However, for
larger values of $p_{T}$ ($>2$ GeV), the interplay between the velocity of sound
and transport coefficient near the QCD phase transition enhances $v_{2}$.
We further point out that Grad's 14 moments approximation
cannot be applied for the calculation of the
one-particle distribution function at the freeze-out.
\end{abstract}

\maketitle

\section{Introduction}

Among many novel discoveries in the studies of relativistic heavy-ion
physics, one of the remarkable facts is the presence of strong collective
flows, which are well described using the ideal hydrodynamic model \cite%
{GeneralHydro}. In this sense, hydrodynamics is expected to be an important
tool to investigate the properties of the QCD matter. If such a scenario is
established, we may further use the hydrodynamic model to infer the initial
condition just after the collision from the final state observables.
However, there are still several open questions in the hydrodynamic
description of relativistic heavy ion collisions \cite{openproblem} and more
extensive studies are necessary. The effect of dissipation is one of such
problems.

The effect of shear viscosity in heavy ion collisions has been investigated
by several authors \cite%
{Romatschke1,Heinzetal,HeinzSong1,HeinzSong2,chaudhuri,pasi, teany}, and it
was found that the elliptic flow parameter $v_{2}$ can be significantly
affected. On the other hand, the effect of bulk viscosity has not been
analyzed in detail yet. Roughly speaking, while the shear viscosity acts as
the resistance against the deformation of a fluid element, the bulk
viscosity acts against the expansion or compression of the fluid. Thus, one
may naturally expect that shear viscosity affects directly $v_{2}$, which
characterizes the spatial anisotropy of the dynamics. However, this does not
mean that the bulk viscosity plays a secondary role. Because of the
anisotropic expansion and compression of the produced matter, the bulk
viscosity should alter the anisotropic flow as well.

Another important factor is the behavior of the bulk viscosity coefficient, $%
\zeta $, near the QCD phase transition temperature $T_{c}$. Recently,
lattice QCD simulations \cite{Karsch,AokiEoS} suggest that $\zeta $ is
strongly enhanced near $T_{c}$, although the shear viscosity coefficient is
suppressed \cite{Nakamura}. A similar enhancement will appear even below $%
T_{c}$, as is shown in \cite{Noronha}, where $\zeta $ in the hadronic phase
increases also quickly towards $T_{c}$. Considering the fact that the fluid
dynamics slows down near $T_{c}$ due to the small sound velocity $c_{s}$,
the critical behavior of $\zeta $ discussed above may play a crucial role on 
$v_{2}$.

As far as we know, the effect of the bulk viscosity on $v_{2}$ was
calculated in \cite{HeinzSongBulk} where the effect of the phase transition
is also considered. As a matter of fact, a quantitative estimate of the
effect of the phase transition in dissipative hydrodynamics is not
straightforward because the above critical behavior of the bulk viscosity
should be consistently constructed with the equation of state (EoS).

As a reliable EoS, the first choice would be the results of the lattice QCD
calculation for the QGP phase and of the hadron resonance gas for the hadron
phase. However, the connection of these two EoS's is not trivial because no
reliable information is available near $T_{c}$. In particular, $c_{s}$ near $%
T_{c}$ is very sensitive to the way that the two phases are connected and it
is important to check how these ambiguities influence $v_{2}$.

The purpose of this paper is to investigate the effect of bulk viscosity on
the elliptic flow within a 2+1 dimensional hydrodynamic modeling,
incorporating the critical behaviors from $c_{s}$ and $\zeta $ near $T_{c}$.
As was pointed out already \cite{HeinzSong1,teany, Romatschke1}, $v_{2}$ is
affected\ also\ by the dissipative corrections on the one-particle
distribution function at the freeze-out. However, we will not consider these
corrections in this work because the 14 moments approximation, which is the
method commonly employed to evaluate this correction, is inapplicable for
the freeze-out values of bulk viscosity, as we will discuss later.

This paper is organized as follows. In Section \ref{sec2}, we briefly review
the equations of causal dissipative hydrodynamics in the hyperbolic
coordinate system and present the SPH (smoothed particle hydrodynamics)
parameterization for the numerical implementation. In Section \ref{sec3}, we
set up the parameters of our calculation, such as the initial condition, the
EoS and the transport coefficients. We futher discuss the freeze-out process
and show the intrinsic problem of the 14 moments approximation. The
numerical results for the effects of bulk viscosity on $v_{2}$ are given in
Section \ref{sec4}. Concluding remarks are given in Section \ref{sec5}.

\section{Relativistic Dissipative Hydrodynamics}

\label{sec2}

\subsection{Memory Function Method in a Hyperbolic Coordinate System}

As is well known, the application of dissipative hydrodynamics in the
relativistic regime requires some caution. Recently, it became clear that
the instability of the hydrodynamic model can be induced by the violation of
relativistic causality \cite{dkkm3}. Thus, the relativistic Navier-Stokes
(Landau-Lifshitz) theory \cite{LL} is essentially unstable and not
applicable for the complex systems generated in the relativistic heavy ion
collisions. Several formulations of relativistic dissipative hydrodynamics
have been proposed so far to solve the problem of acausality and instability 
\cite{MullerReviw}.

In this work, we will use the memory function method \cite{dkkm,dkkm4}. This
theory is formulated in such a way that the magnitude of the bulk viscosity
has naturally a lower bound and we can carry out stable numerical
simulations even for ultra-relativistic initial conditions. Here, we present
the basic equations of this formalism in the hyperbolic coordinate system.

For simplicity, we consider the case of vanishing baryon chemical potential
where only the conservation of the energy and momentum is required. For a
general metric $g_{\mu \nu }$, we have 
\begin{equation}
\frac{1}{\sqrt{-g}}\partial _{\mu }\left( \sqrt{-g}T^{\mu \nu }\right)
+\Gamma _{\lambda \mu }^{\nu }T^{\lambda \mu }=0,  \label{conserv}
\end{equation}%
where $\Gamma _{\mu \lambda }^{\nu }$ is the Christoffel symbol and $g$ is
the determinant of $g_{\mu \nu }$.

We use the Landau definition for the local rest frame and assume, as usual,
that the thermodynamic relations are valid in this frame. In this work, we
further ignore the shear viscosity. Then, the energy-momentum tensor is
expressed as 
\begin{equation}
T^{\mu \nu }=\left( \varepsilon +p+\Pi \right) u^{\mu }u^{\nu }-\left( p+\Pi
\right) g^{\mu \nu },  \label{NeTmunu}
\end{equation}%
where, $\varepsilon $, $p$, $u^{\mu }$ and $\Pi $ are the energy density,
pressure, four velocity and bulk viscosity, respectively.

As was discussed in \cite{dkkm4}, one of the important factors to obtain a
stable causal dissipative hydrodynamics is to consider the deformation of
the fluid element in the hydrodynamic flow. We introduce a reference density 
$\sigma $ defined by%
\begin{equation*}
\frac{\dot{\sigma}}{\sigma }=\theta =u_{;\mu }^{\mu },
\end{equation*}%
where $\theta $ is the expansion rate of a fluid element and $;$ denotes the
covariante derivative. The above relation can also be expressed in the form
of a continuity equation for $\sigma $, 
\begin{equation}
\frac{1}{\sqrt{-g}}\partial _{\mu }\left( \sqrt{-g}\sigma u^{\mu }\right) =0%
\text{.}  \label{volume}
\end{equation}

The equation for the entropy production is written as, 
\begin{equation}
\frac{1}{\sqrt{-g}}\partial _{\mu }\left( \sqrt{-g}su^{\mu }\right) =-\frac{%
\Pi }{T}\frac{1}{\sqrt{-g}}\partial _{\mu }\left( \sqrt{-g}u^{\mu }\right) ,
\label{entrop_conserv}
\end{equation}%
where $s$ is the entropy density. From this, we can define the thermodynamic
force $F$ associated with the bulk viscosity as $\sqrt{-g}^{-1}\partial
_{\mu }\left( \sqrt{-g}u^{\mu }\right) $. In the memory function method, an
irreversible current $J$ is induced by the thermodynamic force as 
\begin{equation}
\frac{J(\tau )}{\sigma (\tau )}=-\int_{\tau _{0}}^{\tau }d\tau ^{\prime }%
\frac{1}{\tau _{R}(\tau ^{\prime })}\exp {\left( -\int_{\tau ^{\prime
}}^{\tau }\frac{d\tau ^{\prime \prime }}{\tau _{R}(\tau ^{\prime \prime })}%
\right) }\frac{F(\tau ^{\prime })}{\sigma (\tau ^{\prime })}+\frac{J(\tau
_{0})}{\sigma (\tau _{0})}\exp {\left( -\int_{\tau _{0}}^{\tau }\frac{d\tau
^{\prime }}{\tau _{R}(\tau ^{\prime })}\right) }.
\end{equation}%
Thus, the equation for the bulk viscosity is obtained by setting $J=\Pi $
and $F=\sqrt{-g}^{-1}\partial _{\mu }\left( \sqrt{-g}u^{\mu }\right) $, and
can be expressed in a differential form as follows 
\begin{equation}
\tau _{\mathrm{R}}u^{\mu }\partial _{\mu }\left( \frac{\Pi }{\sigma }\right)
+\frac{\Pi }{\sigma }=-\frac{\zeta }{\sigma }\frac{1}{\sqrt{-g}}\partial
_{\mu }\left( \sqrt{-g}u^{\mu }\right) ,  \label{GeneralBulk}
\end{equation}%
where $\tau _{\mathrm{R}}$ is the relaxation time.

In hyperbolic coordinates Eqs. (\ref{conserv}), (\ref{entrop_conserv}) and (%
\ref{GeneralBulk}) are explicitly given by 
\begin{eqnarray}
\gamma \frac{d}{d\tau }\left( \frac{\left( \varepsilon +p+\Pi \right) }{%
\sigma }u_{i}\right) &=&\frac{1}{\sigma }\partial _{i}\left( p+\Pi \right) ,
\label{motion} \\
\gamma \frac{d}{d\tau }\left( \frac{s}{\sigma }\right) &=&-\frac{\Pi }{%
T\sigma }\left( \partial _{\mu }u^{\mu }-\frac{\gamma }{\tau }\right) ,
\label{entrop_motion} \\
\tau _{\mathrm{R}}\gamma \frac{d}{d\tau }\Pi +\Pi &=&-\left( \zeta +\tau _{%
\mathrm{R}}\Pi \right) \left( \partial _{\mu }u^{\mu }-\frac{\gamma }{\tau }%
\right) .  \label{Pi}
\end{eqnarray}

In this case $x^{\mu }=(\tau ,\mathbf{r}_{T},\eta )$,where 
\begin{equation}
\tau =\sqrt{t^{2}-z^{2}},~~~\eta =\frac{1}{2}\tanh \left( \frac{t+z}{t-z}%
\right) ,~~~\mathbf{r}_{T}=(x,y),
\end{equation}%
and $\sqrt{-g}=\tau $. The three equations above are solved in the following
calculations.

\subsection{Smoothed Particles Hydrodynamics}

To solve numerically the relativistic hydrodynamic equations, we use the
Smoothed Particle Hydrodynamic (SPH) method \cite{SPH}. See \cite%
{dkkm3,dkkm4,SPH-heavy ion} for the application to the heavy ion dynamics.
For the sake of book keeping, here we just write down explicitly the SPH
equations in hyperbolic coordinates leaving the detailed derivation to the
references above.

In the SPH method, all the hydrodynamic variables are expressed in terms of
a discrete set of Lagrangian coordinates $\left\{ \mathbf{r}_{\alpha }(t),\
\alpha =1,...,N_{SPH}\right\}$ together with a normalized kernel function $%
W\left( \mathbf{r},h\right)$ for the smoothing procedure. The parameter $h$
represents the width of $W$ and serves as a cut-off parameter for short
wavelength modes. For hyperbolic coordinates, this kernel function takes the
form 
\begin{equation}
W\left( \mathbf{\tilde{r}},h\right) =W\left( \eta ,h_{\eta }\right) W\left( 
\mathbf{r}_{T},h_{T}\right) ,
\end{equation}%
with 
\begin{eqnarray}
\int W\left( \eta ,h_{\eta }\right) d\eta &\mathbf{=}&\mathbf{1,} \\
\int W\left( \mathbf{r}_{T},h_{T}\right) d\mathbf{r}_{T} &\mathbf{=}&\mathbf{%
1,}
\end{eqnarray}%
where the two width parameters, $h_{\eta }$ and $h_{T}$, are introduced
separately for $\eta $ and $\mathbf{r}_{T}$, respectively.

The basic procedure of the SPH scheme is to express a reference density $%
\sigma ^{\ast }=\tau \gamma \sigma $ and its current $\mathbf{j=}\sigma
^{\ast }\mathbf{v}$\textbf{,} where $\gamma =1/\sqrt{1-\mathbf{v}%
_{T}^{2}-\tau ^{2}v_{\eta }^{2}}$ is the Lorentz factor, as%
\begin{gather}
\tau \gamma \sigma \rightarrow \sigma _{SPH}^{\ast }\left( \mathbf{r}%
,t\right) =\sum_{\alpha =1}^{N_{SPH}}\nu _{\alpha }W(\mathbf{r}-\mathbf{r}%
_{\alpha }(t),h),  \label{sigma} \\
\mathbf{j}\rightarrow \mathbf{j}_{SPH}\left( \mathbf{r},t\right)
=\sum_{\alpha =1}^{N_{SPH}}\nu _{\alpha }\frac{d\mathbf{r}_{\alpha }(t)}{dt}%
W(\mathbf{r}-\mathbf{r}_{\alpha }(t),h),  \label{current}
\end{gather}%
so that the continuity equation (\ref{volume}) is automatically satisfied 
\cite{SPH-heavy ion}. Eqs. (\ref{sigma}) and (\ref{current}) can be
interpreted as if there are particles associated with each Lagrangian
coordinate $\mathbf{r}_{\alpha }(t)$, with velocity $\mathbf{v}_{\alpha
}(t)=d\mathbf{r}_{\alpha }/dt$, carrying a quantity $\nu _{\alpha }$ for the
reference density $\sigma _{SPH}^{\ast }$. These particles are called SPH
particles. The following calculation is independent of the choice of $\nu
_{\alpha }$ and, hence, we set $\nu _{\alpha }=1$.

The dynamic variables in the SPH scheme are then,%
\begin{equation}
\left\{ \mathbf{r}_{\alpha },\mathbf{u}_{\alpha },\left( \frac{s}{\sigma }%
\right) _{\alpha },\left( \frac{\Pi }{\sigma }\right) _{\alpha };\ \alpha
=1,..,N_{SPH}\right\} ,
\end{equation}%
where they represent, respectively, position, velocity, entropy, and bulk
viscosity associated with the $\alpha $-th SPH particle. The time evolution
of these variables can be obtained from Eqs. (\ref{motion}), (\ref%
{entrop_motion}) and (\ref{Pi}) by expressing them in the SPH
representation. Writing the entropy density $s$ and the bulk viscosity $\Pi $
as 
\begin{eqnarray}
\tau \gamma s &\rightarrow &s_{SPH}^{\ast }\left( \mathbf{r},t\right)
=\sum_{\alpha }\nu _{\alpha }\left( \frac{s}{\sigma }\right) _{\alpha }W(%
\mathbf{r}-\mathbf{r}_{\alpha }(t),h), \\
\Pi &\rightarrow &\Pi _{SPH}=\sum_{\alpha }\nu _{\alpha }\frac{1}{\gamma
_{\alpha }\tau }\left( \frac{\Pi }{\sigma }\right) _{\alpha }W(\mathbf{r}-%
\mathbf{r}_{\alpha }(t),h),
\end{eqnarray}%
we obtain the equation for the bulk viscosity of the $\alpha $-th SPH
particle 
\begin{equation}
\tau _{R_{\alpha }}\gamma _{\alpha }\frac{d}{d\tau }\left( \frac{\Pi }{%
\sigma }\right) _{\alpha }+\left( \frac{\Pi }{\sigma }\right) _{\alpha }=-%
\frac{\zeta _{\alpha }}{\sigma _{\alpha }}\left( \partial _{\mu }u^{\mu }-%
\frac{\gamma }{\tau }\right) _{\alpha },
\end{equation}%
and the equation of motion Eq. (\ref{motion}) \cite{SPH-heavy ion, dkkm3} 
\begin{equation}
\sigma _{SPH}^{\ast }\frac{d}{d\tau }\left( \frac{\left( \varepsilon +p+\Pi
\right) _{\alpha }}{\sigma _{\alpha }}u_{i\ \alpha }\right) =\tau
\sum_{\beta =1}^{N_{SPH}}\nu _{\beta }\sigma _{\alpha }^{\ast }\left( \frac{%
p_{\beta }+\Pi _{\beta }}{\left( \sigma _{\beta }^{\ast }\right) ^{2}}+\frac{%
p_{\alpha }+\Pi _{\alpha }}{\left( \sigma _{\alpha }^{\ast }\right) ^{2}}%
\right) \ \partial _{i}W(|\mathbf{r}_{\alpha }-\mathbf{r}_{\beta }(t)|).
\label{MotionSPH}
\end{equation}%
The r.h.s. of Eq.(\ref{MotionSPH}) is nothing but the SPH representation of
the gradients of pressure and bulk viscosity. We remark that the l.h.s. of
Eq.(\ref{MotionSPH}) contains an implicit velocity dependence through the
Lorentz factor $\gamma $ in the thermodynamic variables. After some
straightforward manipulations, we get 
\begin{equation}
M_{i}^{m}\frac{du_{m}}{d\tau }=F_{i},
\end{equation}%
with%
\begin{eqnarray}
M_{i}^{j} &=&\gamma C\delta _{i}^{m}-\frac{A}{\gamma }g^{lm}u_{i}u_{l}, \\
F_{i} &=&Bu_{i}+\partial _{i}\left( p+\Pi \right) ,
\end{eqnarray}%
and%
\begin{eqnarray}
A &=&\varepsilon +p-\frac{dw}{ds}\left( \frac{\Pi }{T}+s\right) -\frac{\zeta 
}{\tau _{R}}, \\
B &=&\left( \frac{\gamma }{\sigma ^{\ast }}\frac{d\sigma ^{\ast }}{d\tau }-%
\frac{\gamma }{\tau }+\frac{1}{2\gamma }u_{l}u_{m}\frac{dg^{lm}}{d\tau }%
\right) A+\frac{\Pi }{\tau _{R}}, \\
C &=&\varepsilon +p+\Pi .
\end{eqnarray}%
In the above equations, we calculated the four divergence of the velocity as 
\begin{equation*}
\partial ^{\mu }u_{\mu }=-\frac{\gamma _{\alpha }}{\sigma ^{\ast }}\frac{%
d\sigma ^{\ast }}{d\tau }-\frac{g^{ij}u_{i}}{\gamma }\frac{du_{j}}{d\tau }-%
\frac{u_{i}u_{j}}{2\gamma }\frac{dg^{ij}}{d\tau },
\end{equation*}%
and, from the continuity equation, 
\begin{equation*}
\frac{d\sigma ^{\ast }}{dt}=\frac{1}{\sigma ^{\ast }}\ \mathbf{j}_{SPH}\cdot
\nabla \sigma _{SPH}^{\ast }-\nabla \cdot \mathbf{j}_{SPH}.
\end{equation*}

As mentioned in the introduction, we study the transverse dynamics near the
central rapidity region, assuming the Bjorken scaling behavior in $\eta$,
that is, $u_{\eta }=0$. Then the dynamics is restricted on the transverse
plane and is simplified as 
\begin{equation}
\left( \gamma C\delta ^{ij}+\frac{A}{\gamma }u_{T}^{i}u_{T}^{j}\right) \frac{%
du_{T}^{j}}{d\tau }=Bu_{T}^{i}-\partial _{i}\left( p+\Pi \right) ,
\end{equation}%
with%
\begin{equation}
B=\left( \frac{\gamma }{\sigma ^{\ast }}\frac{d\sigma ^{\ast }}{d\tau }-%
\frac{\gamma }{\tau }\right) A+\frac{\Pi }{\tau _{R}}.
\end{equation}%
These are the SPH representations of our equations which will be solved.

\section{Parameter Settings of the Model}

\label{sec3}

\subsection{Initial Conditions}

One of the central issues of hydrodynamical simulations of relativistic
heavy-ion collisions is to infer what are the real initial conditions
attained in the instant of the collision. Here we just assume the initial
conditions suggested by \cite{HiranoIC,HeinzSong1} for the sake of
comparison. The initial energy density is parameterized as 
\begin{eqnarray}
\varepsilon _{0} &=&K\left\{ T_{A}\left( x+\frac{b}{2},y\right) \left[
1-\left( 1-\frac{\sigma _{NN}T_{B}\left( x-\frac{b}{2},y\right) }{B}\right)
^{B}\right] \right.  \notag \\
&&\left. +T_{B}\left( x-\frac{b}{2},y\right) \left[ 1-\left( 1-\frac{\sigma
_{NN}T_{A}\left( x+\frac{b}{2},y\right) }{A}\right) ^{A}\right] \right\} ,
\end{eqnarray}%
where $A=B=197$ for Au-Au collisions with the nucleon-nucleon cross section $%
\sigma _{NN}=40$ $\mathrm{mb}$. The free parameter $K$ is fixed so that the
maximum energy density $\varepsilon _{0}$ is $30$ GeV for central
collisions, as is done in \cite{HeinzSong1}. The term $T_{A}\left(
x,y\right) $ is the nuclear thickness function defined as 
\begin{equation}
T_{A}\left( x,y\right) =\int dz\frac{\rho _{0}}{1+\exp \left[ \left(
r-R_{A}\right) /\lambda \right] },
\end{equation}%
where $\rho _{0}=0.17$ $\mathrm{fm}^{-3}$, $R_{Au}=6.37$ $\mathrm{fm}$ and $%
\lambda =0.54$ $\mathrm{fm}$, respectively. The impact parameter is chosen
as $b=7$ $\mathrm{fm}$.

\subsection{Equation of State}

\begin{figure}[tbp]
\centering \includegraphics[scale=0.25]{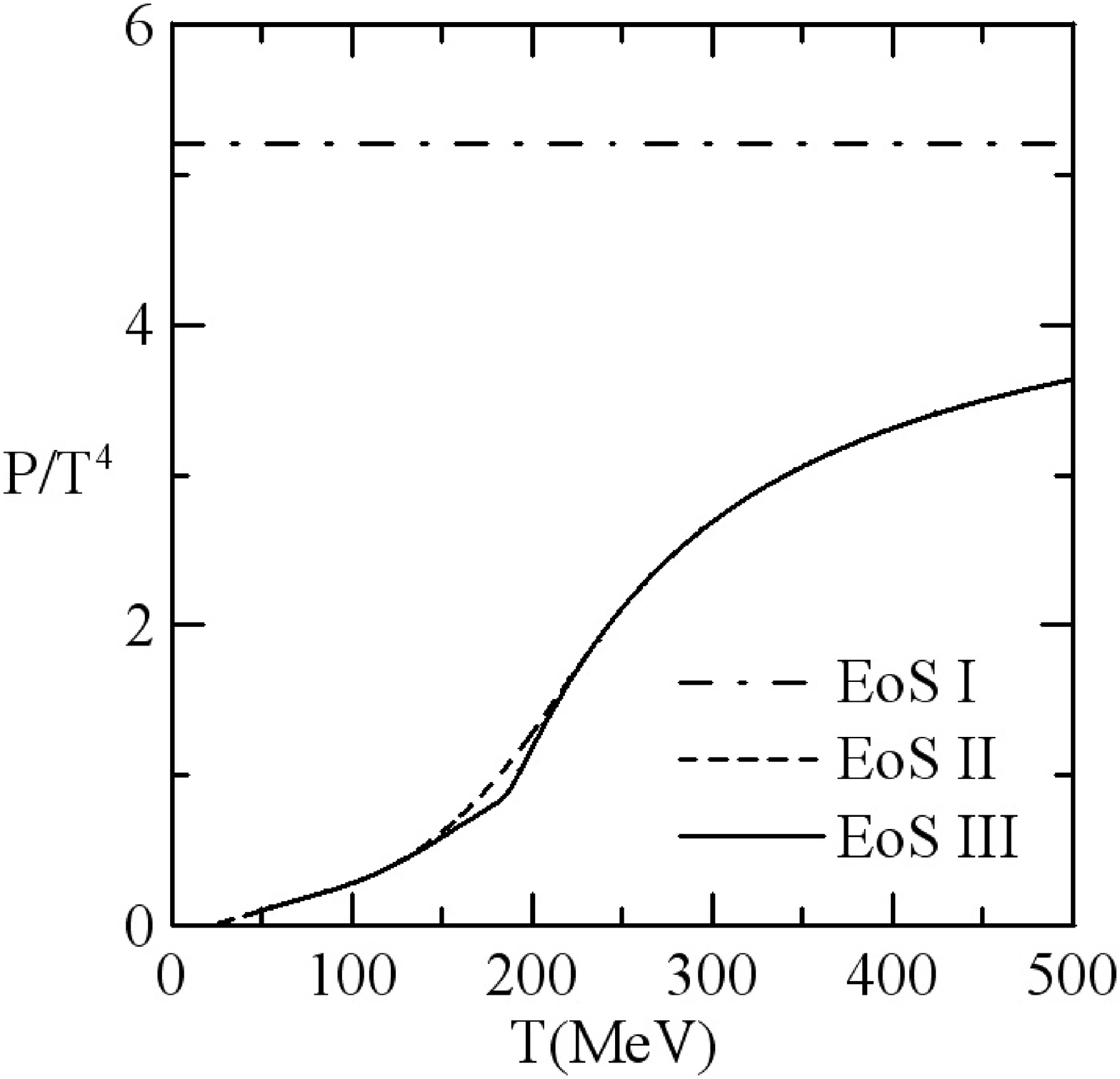}
\caption{The temperature dependence of the EoS. The dotted and solid lines
represent the EoS's with the QCD phase transition. The dot-dashed line
denotes the ideal gas of massless quarks EoS.}
\label{eos}
\end{figure}

We consider three different EoS's; one for an ideal gas of massless quarks
and the other two corresponding to different connections between the QGP and
hadron gas phases, as shown in Fig. \ref{eos}. The dash-dotted line
corresponds to the ideal massless quark gas, which we refer to as EoS I. The
dotted line (EoS II) and solid line (EoS III) are both constructed by
connecting the EoS from lattice QCD calculations with three flavors for the
QGP phase \cite{AokiEoS} and that of the hadron resonance gas for the hadron
phase. However, as was mentioned in the introduction, it is not well
established how the lattice EoS is connected to that of the hadron resonance
gas EoS. In EoS II the two phases are connected smoothly for a wide
interpolation domain of the temperature, $100\lessapprox T\lessapprox 290$
MeV, whereas in EoS III the connetion is done within a narrower
interpolation domain, $180\lessapprox T\lessapprox 210$ MeV (see \cite%
{ConnectionEoS} for the connection method). Thus, compared with the EoS II,
the EoS III is more faithful, both to the lattice QCD and the hadronic
resonance gas values except in a narrow transition region near $T\simeq 200$
MeV. On the other hand, since the connection is rather abrupt, a kink-like
structure appears at the QCD phase transition. 

\begin{figure}[tbp]
\centering \includegraphics[scale=0.25]{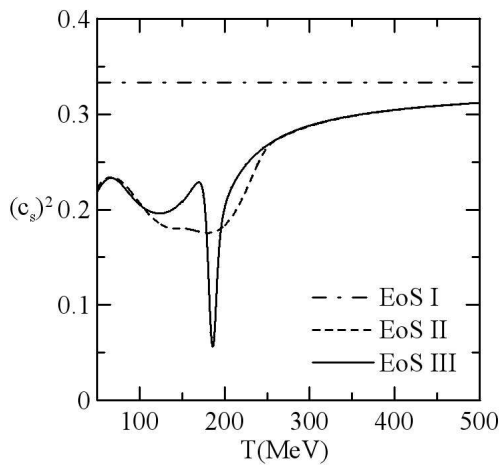}
\caption{The temperature dependence of $c^2_s$. The dotted and solid lines 
represent EoS II and EoS III, respectively. The dot-dashed line
denotes the ideal gas of massless quarks EoS.}
\label{sound}
\end{figure}

The temperature dependence of the pressure and $c_{s}$ are shown in Figs. %
\ref{eos} and \ref{sound}, respectively. One can see that, although the
behavior of the pressure is not drastically modified, $c_{s}$ is still\ very
sensitive to the choice of connection; $c_{s}$ in EoS III has a narrow
minimum reaching a much smaller value than that of EoS II. The nature of the
transition of EoS III is almost that of the first order phase transition. If
the transition were of first order, $c_{s}$ would vanish at the phase
transition. Since the acceleration of the fluid is directly related to $%
c_{s} $, this difference would affect significantly the hydrodynamic
evolution of the system near the phase transition. Thus, we perform a
comparative study of the behavior of $v_{2}$ calculated using these
equations of state.

\subsection{Transport Coefficients}

\begin{figure}[tbp]
\centering \includegraphics[scale=0.25]{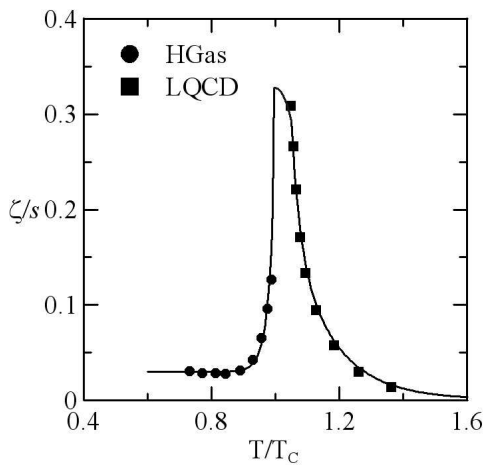}
\caption{The fit for $\protect\zeta/s$. The circles and squares denote the
result obtained from Lattice QCD calculation and the hadron
resonance gas, respectively.}
\label{zeta}
\end{figure}

The behavior of the transport coefficients $\zeta $ and $\tau _{R}$ has not
yet been established. There are several approaches to calculate transport
coefficients. A well-known method is to use the Green-Kubo-Nakano (GKN)
formula. Exactly speaking, the validity of the GKN formula is, however, not
clear in the relativistic regime since it was developed for Newtonian
fluids. Recently it has been shown that even in the presence of memory
effects, the viscosity coefficients coincide with those of the GKN formula,
at least, in the leading order approximation \cite{knk}. Here, we use the
result from lattice QCD \cite{Karsch}, calculated with the GKN formula, for
the QGP phase. For the hadron phase, we use the result obtained recently by 
\cite{Noronha}. Then, the bulk viscosity coefficient shows a maximum at $%
T_{c}$ and starts to decrease almost exponentially as the system departs
from this critical region. The coefficient vanishes in the high temperature
limit in the QGP phase whereas in the low temperature limit (hadronic gas),
it converges to a finite value. We fitted these results by analytic
expressions and connected them by a parabola around $T_{c}=200$ MeV. The
result is%
\begin{equation*}
\frac{\zeta }{s}=\left\{ 
\begin{array}{ll}
A_{1}x^{2}+A_{2}x-A_{3} & (0.995T_{c}\geq T\geq 1.05T_{c}) \\ 
\lambda _{1}exp(-(x-1)/\sigma _{1})+\lambda _{2}exp(-(x-1)/\sigma _{2})+0.001
& (T>1.05T_{C}) \\ 
\lambda _{3}exp((x-1)/\sigma _{3})+\lambda _{4}exp((x-1)/\sigma _{4})+0.03 & 
(T<0.995T_{C}),%
\end{array}%
\right.
\end{equation*}%
where $x=T/T_{C}$. The fitted parameters are $\lambda _{1}=\lambda _{3}=0.9$%
, $\lambda _{2}=0.25$, $\lambda _{4}=0.22$, $\sigma _{1}=0.025$, $\sigma
_{2}=0.13,$ $\sigma _{3}=0.0025$, $\sigma _{4}=0.022,$ $A_{1}=-13.77$, $%
A_{2}=27.55$ and $A_{3}=13.45$. The Fig. \ref{zeta} shows this fit and
compares it with the results from \cite{Karsch,Noronha}.

Inspired by the result for the relaxation time of the causal shear viscosity
coefficient \cite{knk}, we use the following parameterization for the
relaxation time of the bulk viscosity, 
\begin{equation}
\tau _{R}=\frac{\zeta }{\epsilon +p}b_{\Pi }.
\end{equation}

\begin{figure}[tbp]
\centering \includegraphics[scale=0.25]{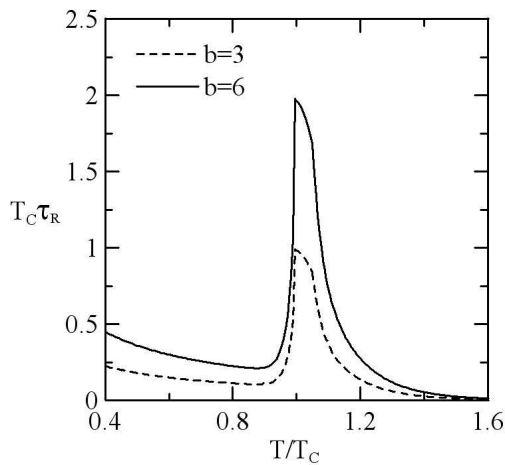}
\caption{The temperature dependence of the bulk viscosity relaxation time. The solid and dotted
lines represent $b_{\Pi }=6$ and $b_{\Pi }=3$, respectively.}
\label{tau}
\end{figure}

In principle, the parameter $b_{\Pi }$ is a function of thermodynamical
quantities constrained by the causality condition \cite{dkkm4}. Here, for
simplicity, we will always consider $b_{\Pi }$ as a constant. Due to this
assumption, the relaxation time also shows critical behavior as in the case
of the bulk viscosity coefficient, as is shown in Fig. \ref{tau}.

\subsection{Freeze out and Kinetic Corrections}

From extensive studies of the shear viscosity, it is known that there are
two different origins for the dissipative corrections to the elliptic flow.
One is the effect of viscosity on the collective evolution of the fluid
itself. The other is the correction to the one-particle distribution
function at the freeze-out hyper surface. It was shown that the latter
correction is more important than the former for the shear viscosity \cite%
{HeinzSong1,teany}.

The correction to the one-particle distribution function is usually
estimated using Grad's 14 moments approximation. However, the applicability
of this approach to the heavy-ion scenario is not so obvious. In the
following, we will show that this procedure exhibits intrinsic problems at
least for the bulk viscosity.

In the 14 moments approximation, the one-particle distribution function for
bosons is expressed as 
\begin{equation}
f=f_{0}+\delta f\approx f_{0}+f_{0}(1+f_{0})(E_{0}+D_{0}(p^{\mu }u_{\mu
})+B_{0}(p^{2}-4(p^{\mu }u_{\mu })^{2}))\Pi ,
\end{equation}%
where $f_{0}=f_{0}\left( p\right) $ refers to the Bose-Einstein distribution
function and $p^{\mu }$ is the four momentum of a particle. In this
expression, we neglected the contributions from the shear viscosity and the
heat conduction. See Appendix \ref{app:a} for details \footnote{%
The correction to the one-particle distribution function due to the bulk
viscosity is also discussed in \cite{teany}. Their expression, however, does
not include the scalar and linear terms in momentum to the correction of the
distribution function, that is, they neglect terms like $E_{0}$ and $%
D_{0}(p^{\mu }u_{\mu })$. However, this form of distribution function breaks
the Landau conditions and satisfies only one of the normalization conditions, Eqs. (%
\ref{norm_1}) and (\ref{norm_2}).}.

\begin{figure}[tbp]
\centering \includegraphics[scale=0.25]{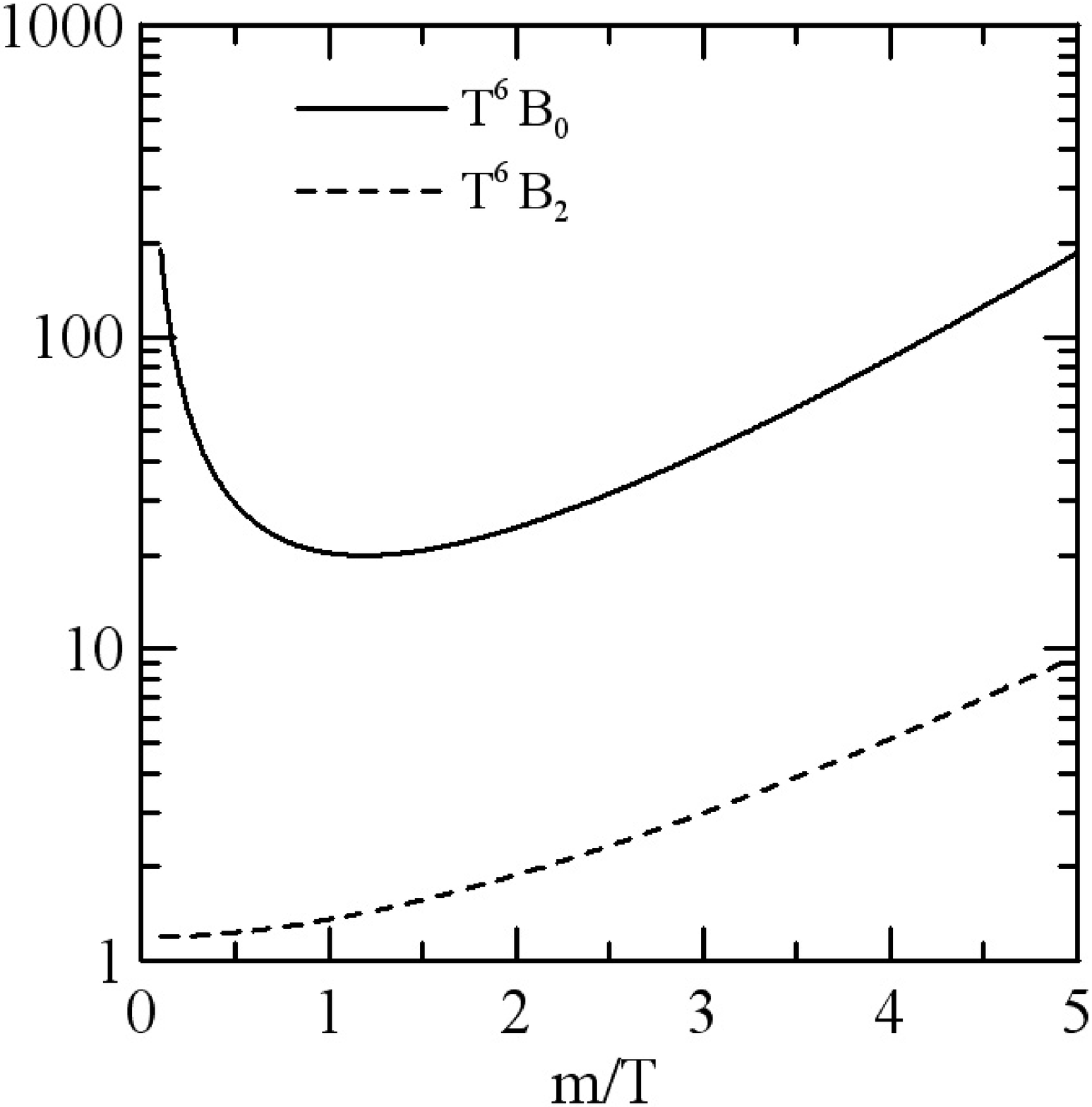}
\caption{The coefficients $B_{0}$ and $B_{2}$ as a function of $m$/$T$. See
Appendix \protect\ref{app:a} for details.}
\label{BB}
\end{figure}

\begin{figure}[tbp]
\centering \includegraphics[scale=0.25]{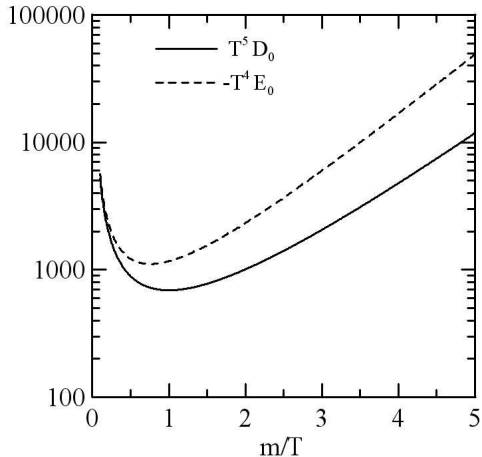}
\caption{The coefficients $D_{0}$ and $E_{0}$ as a function of $m$/$T$. See
Appendix \protect\ref{app:a} for details.}
\label{DE}
\end{figure}

\begin{figure}[tbp]
\centering \includegraphics[scale=0.25]{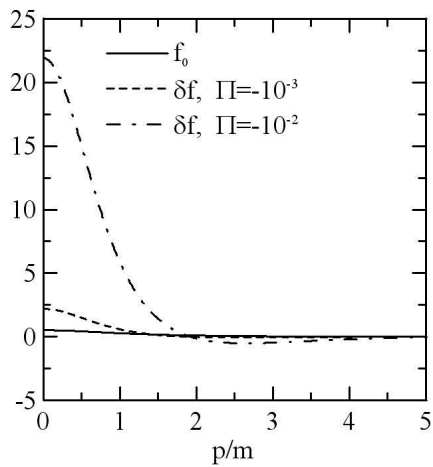}
\caption{The correction to the one-particle distribution function for
different values of the bulk viscosity $\Pi$. The solid line represents $f_0$
and the dotted and dash-dotted lines represent $\protect\delta f$ for $\Pi$ $%
=$ $10^{-3}$and $10^{-2}$fm$^{-4}$, respectively.}
\label{dist}
\end{figure}

In Figs. \ref{BB} and \ref{DE}, the coefficients of the viscous corrections $%
B_{0}$, $D_{0}$ and $E_{0}$ are shown as a function of $m/T$. For the sake
of comparison, we plot also $B_{2}$ in Fig. \ref{BB}, which is the
coefficient corresponding to the viscous correction from the shear viscosity
(see Eq. (\ref{lastname})). One can see that the bulk correction $B_{0}$ is
one order of magnitude larger than the shear correction $B_{2}$. By using
these coefficients with $m=140$ MeV and $T=130$ MeV, we calculated the
correction to the one-particle distribution function for different values of
the bulk viscosity, which are shown in Fig. \ref{dist}. In our simulations,
the typical value of the bulk viscosity at the freeze-out hyper surface
reaches $\Pi \approx 10^{-2}\ $fm$^{-4}$, which means that $\delta f$ is
much bigger than $f_{0}$ itself. Then, the corrected one-particle
distribution function $f$ can be even negative. The negative corrections
come from the normalization conditions, Eqs. (\ref{cond4}) and (\ref{cond5}%
), which imply that 
\begin{eqnarray}
{\displaystyle\int }\frac{d^{3}\vec{K}}{\left( 2\pi \right) ^{3}}\delta f
&=&0,  \label{norm_1} \\
{\displaystyle\int }\frac{d^{3}\vec{K}}{\left( 2\pi \right) ^{3}}E\delta f
&=&0.  \label{norm_2}
\end{eqnarray}

Obviously, the 14 moments approximation is consistent only when $\delta f$
is much smaller than the equilibrium distribution function $f_{0}$. That is,
the magnitude of bulk viscosity $\Pi $ should be, at least, smaller than $%
10^{-4}$ fm$^{-4}$ at the freeze-out for pions. This value of bulk viscosity
is very small and corresponds only to $\sim 0.05\%$ of the value of pressure
of a hadron resonance gas at $130$ MeV. Usually, the procedure of freeze-out
in hydrodynamics does not impose any constraint on the magnitude of the bulk
viscosity and it is not clear that such small values of viscosity will be
achieved. For example, in our simulations, the value of the ratio between
the bulk viscosity and the pressure is around $10\%$, which seems to be a
small correction but is already too large for the 14 moments approximation.
This situation is unchanged even for heavier particles ($m/T>1$).

One may notice that Grad's method was originally formulated for a single
component fluid and, to discuss the appropriate correction to the
one-particle distribution function, we have to generalize it to a multiple
component fluid. Thus, one might argue that if we consider the
multi-component fluid the limitation discussed above might be relaxed. In
this case, each hadronic species can have its own bulk viscosity and
pressure, $\Pi _{(i)}$ and $p_{(i)}$, where%
\begin{eqnarray}
\Pi &=&\sum_{i}\Pi _{(i)},  \label{bulk_mult1} \\
p &=&\sum_{i}p_{(i)}.
\end{eqnarray}%
Here the index $(i)$\ represents the different hadronic species. In the
context of the first order theory, the ratio $\Pi _{(i)}/\Pi $ should be
proportional to the ratio of the coefficients $\zeta _{(i)}/\zeta $, that
is, 
\begin{equation}
\frac{\Pi _{(i)}}{\Pi }=\frac{\zeta _{(i)}}{\zeta }=\frac{s_{(i)}}{s}\sim
10^{-1}.  \label{bulk_mult2}
\end{equation}%
In the above expression, $s$\ is the entropy density and $s_{(i)}$\ is the
entropy density contribution from the $i$-th\ specie.\textbf{\ }Then one
should use the ratio for each species, $\Pi _{(i)}/p_{(i)}$, to estimate the
dissipative correction. From the discussion above, the magnitude of the bulk
viscosity for pions, $\Pi _{(\pi )}$, should be limited to be smaller than $%
10^{-4}$ fm$^{-4}$. From this, we conclude that $\Pi _{(\pi )}/p_{(\pi )}$
at $T=130$ MeV should be at maximum $2\%$, instead of $0.05\%$, for the $14$
moment method to be applicable. Even though the limit of applicability is
relaxed, it is still not clear that such a constraint will always be
satisfied in a realistic hydrodynamic scenario. Furthermore, it should be
noted that the generalization of the 14 moments method to a multiple
component fluid is not trivial, neither unique. There are several different
solutions proposed \cite{PPV,hirano}.

We conclude that the applicability of the 14 moments approximation is not
clear in heavy-ion collisions. For this reason, we will not consider this
correction to the single-particle distribution function in the following
discussions.

\section{Numerical results}

\label{sec4}

We calculate the elliptic flow, $v_{2}=\left\langle \cos (2\phi
)\right\rangle $, for pions as a function of the transverse momentum with a
freeze out temperature of $130$ MeV by the Cooper-Frye method \cite%
{SPH-heavy ion}. Unless we say otherwise, we take $b_{\Pi }=6$. Results for
constant values of the ratio $\zeta $/$s$ will be shown for the sake of
comparison.

\begin{figure}[tbp]
\centering \includegraphics[scale=0.25]{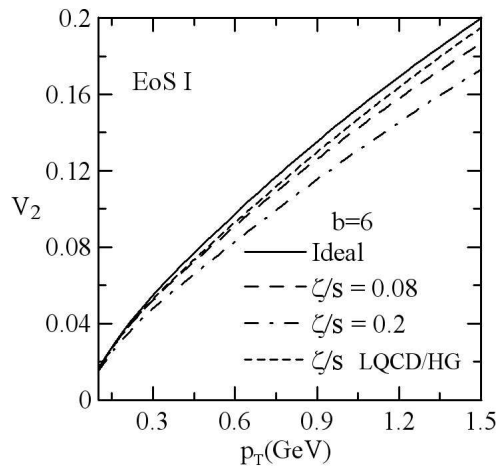}
\caption{$v_2$ as a function of $p_{T}$ for EoS I with different values of $\protect%
\zeta/s$. The solid line corresponds to the ideal result, the dashed and
dash-dotted lines corresponds respectively to $\protect\zeta /s=$ $0.08$ and 
$0.2$ and the dotted line is the result for $\protect\zeta /s$ with critical
behavior.}
\label{Qgas}
\end{figure}

\begin{figure}[tbp]
\centering \includegraphics[scale=0.25]{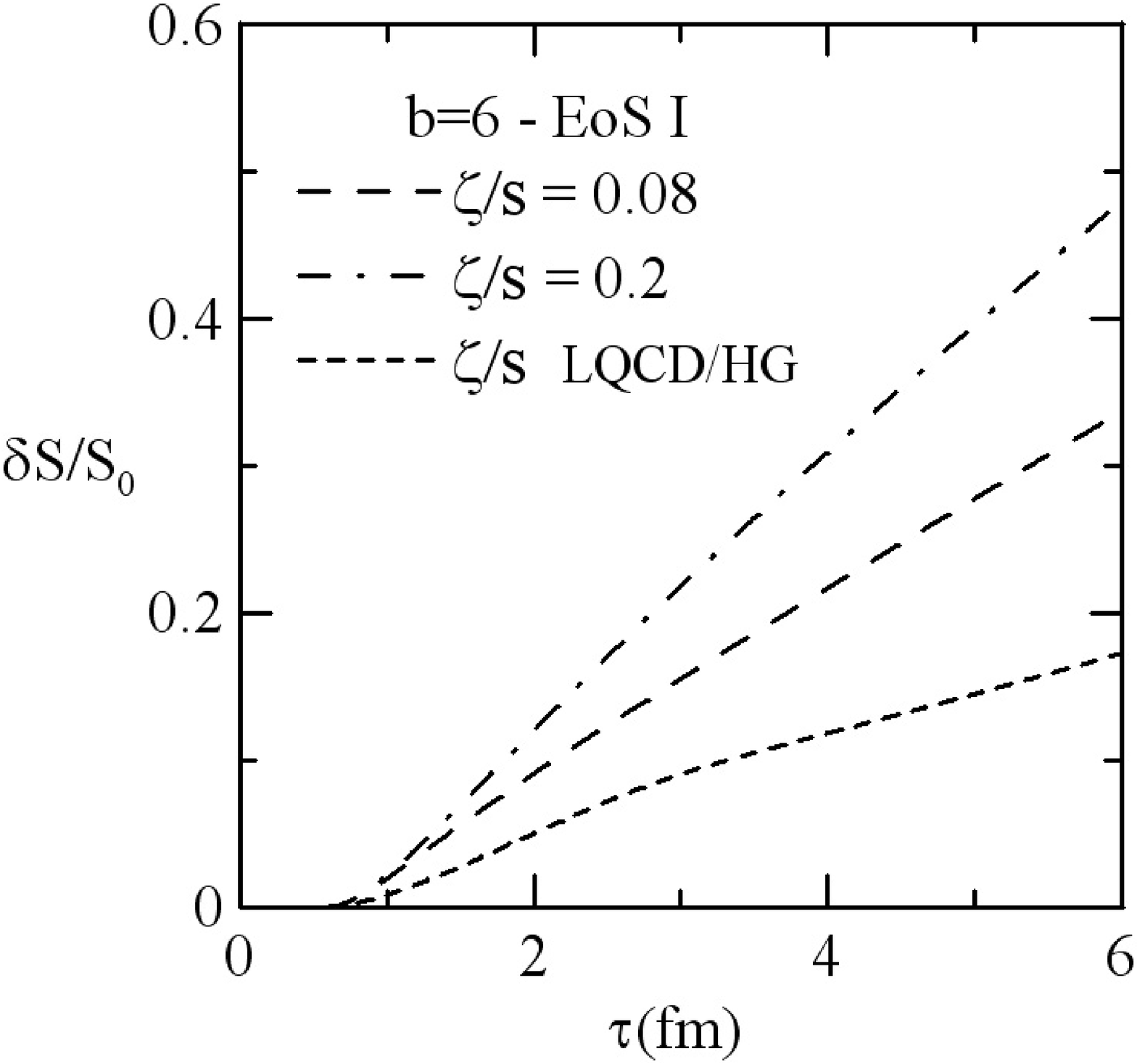}
\caption{The time evolution of the entropy production for different values of $\protect%
\zeta/s$ with EoS I (in units of the initial entropy). The dashed and
dash-dotted lines corresponds respectively to $\protect\zeta /s=$ $0.08$ and 
$0.2$ and the dotted line is the result for $\protect\zeta /s$ with
critical behavior.}
\label{EntropyProd}
\end{figure}

In Fig. \ref{Qgas}, we show $v_{2}$ calculated with the ideal gas of
massless quarks EoS (EoS I). Analogous to the case of the shear viscosity, $%
v_{2}$ decreases as $\zeta $ increases, which is expected for smooth
expanding cases, since the viscosity converts a part of the collective
kinetic energy to heat. Even when we use the $\zeta $/$s$ with the critical
behavior of the QCD phase transition, this surppression of $v_{2}$ is
essentially the same, as shown by the dotted line. As a matter of fact, we
found a direct correlation between $v_{2}$ and the entropy production. In
Fig. \ref{EntropyProd}, the entropy production is shown for different values
of $\zeta $/$s$, and one can notice that the larger the entropy production,
the bigger the suppression of $v_{2}$.

\begin{figure}[tbp]
\centering \includegraphics[scale=0.25]{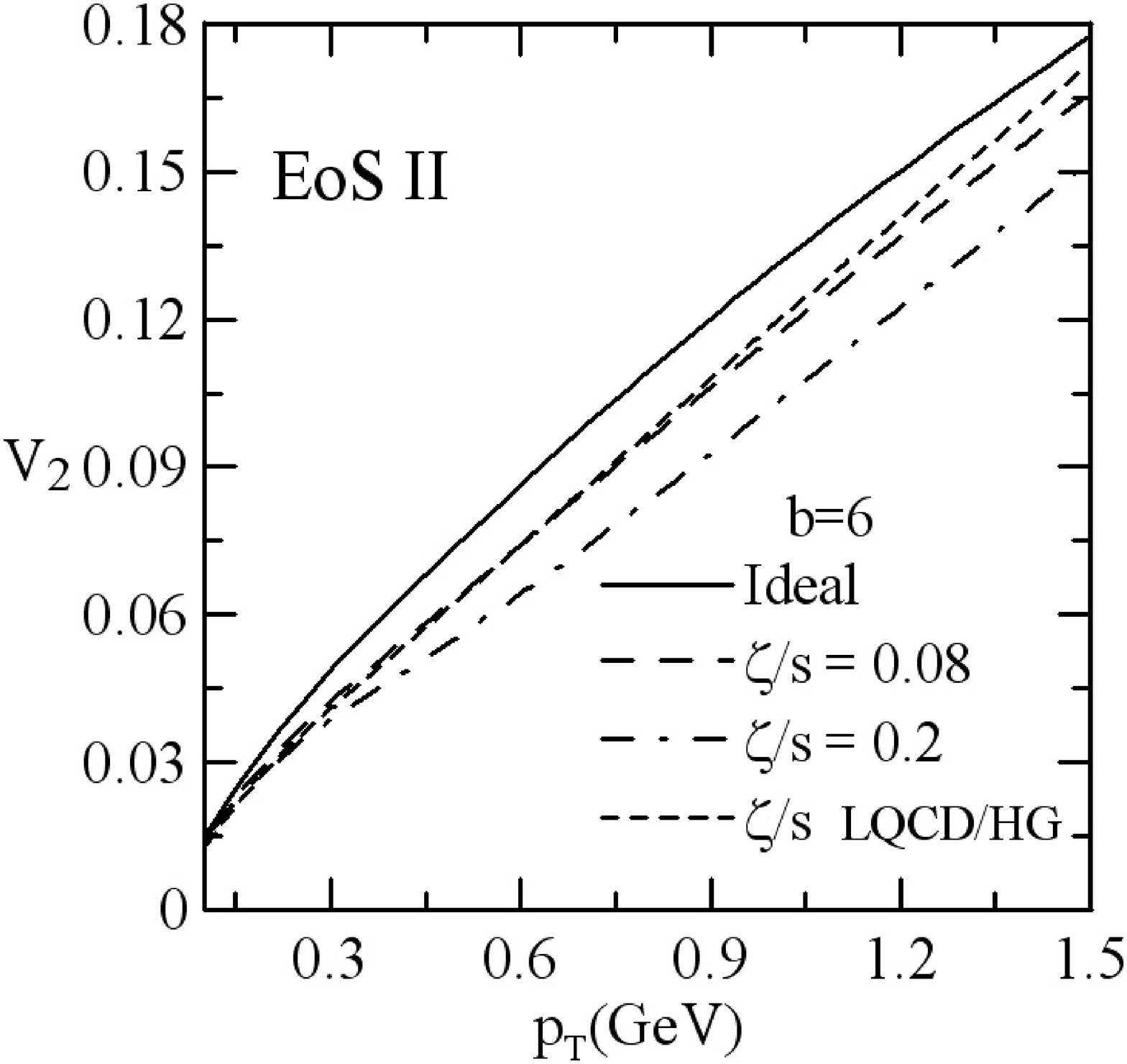}
\caption{$v_{2}$ as a function of $p_{T}$ with EoS II. The figure shows $v_{2}$ for different values of $\protect\zeta /s$, at low transverse momentum.
The solid line correspond to the ideal result, the dashed and dash-dotted
lines corresponds respectively to $\protect\zeta /s=$ $0.08$ and $0.2$, and
the dotted line is the result for $\protect\zeta /s$ with critical behavior.}
\label{Lattice}
\end{figure}

Next, we consider the effect of the QCD phase transition through the EoS. In
Fig. \ref{Lattice}, $v_{2}$ of viscous (dashed, dash-dotted and dotted
lines) and ideal (solid line) fluids are shown for the EoS II. The dashed
and dash-dotted lines correspond to constant $\zeta $/$s$ cases, and the
dotted line represents the $\zeta /s$ with the QCD phase transition.
Although the general behavior is similar to the case of EoS I, the
curvatures of $v_{2}$ for the corresponding cases are changed: $v_{2}$ is
reduced at low $p_{T}$ but starts to increase at high $p_{T}$. This effect
at high $p_{T}$ is enhanced when $\zeta /s$ exhibits the critical behavior.

\begin{figure}[tbp]
\centering \includegraphics[scale=0.25]{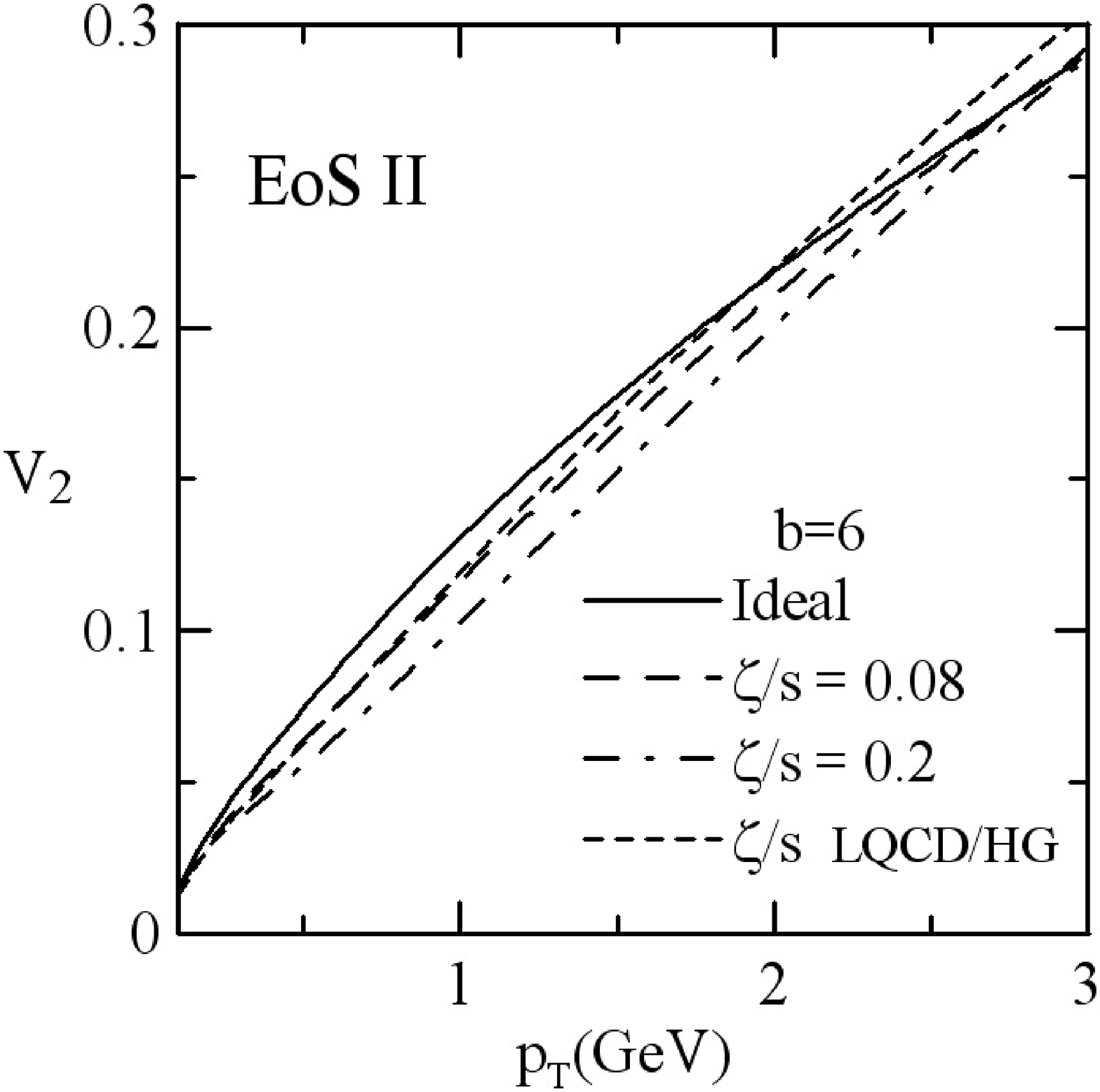}
\caption{$v_{2}$ as a function of $p_{T}$ with EoS II. The figure shows the
behaviors for different values of $\protect\zeta /s$, at high transverse momentum.
The solid line corresponds to the ideal result, the dashed and dash-dotted
lines correspond respectively to $\protect\zeta /s=$ $0.08$ and $0.2$ and
the dotted line is the result for $\protect\zeta /s$ with critical behavior.}
\label{LatticeLarge}
\end{figure}

To see this effect in detail, we show $v_{2}$ up to $p_{T}=$ 3 GeV in Fig. %
\ref{LatticeLarge}. For $p_{T}>1.5$ GeV, $v_{2}$ starts to increase and
eventually surpasses the value of the ideal case. No significant difference
was observed for the EoS III compared to the case of EoS II, although this
EoS displays a sharper transition from the QGP phase to the hadron phase. As
discussed below, this interesting feature is the result of an interplay
between the viscosity and the sound velocity. 

\begin{figure}[tbp]
\centering \includegraphics[scale=0.3]{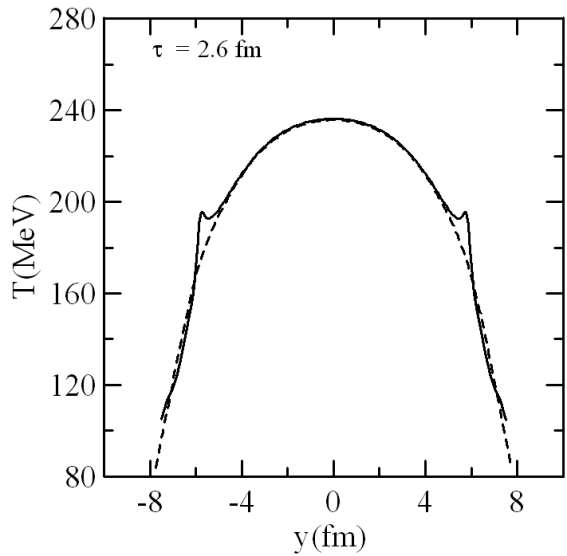}
\caption{Comparison of the temperature profiles for the ideal and viscous
fluids. The viscous fluid, shown by the solid line, shows an accumulation of
matter near the phase transition due to the effect of viscosity.}
\label{shockwave}
\end{figure}

First, note that the enhancement of $v_{2}$ at high $p_{T}$ is not observed
for the EoS I case. Thus, this effect should be attributed to the presence
of the QCD phase transition (or cross over). In Fig. \ref{shockwave}, the
temperature distribution calculated with EoS II at $\tau =2.6$ fm is shown.
The solid and dashed lines represent the ideal and viscous results,
respectively. One can easily see that, for the viscous case, the behavior of
the fluid flow is not monotone in the region where $c_{s}$ has its minimum.
This is because the flow of the internal matter, which has a higher
temperature, tends to catch up the foregoing fluid elements, generating a
non-monotonous velocity field in the radial direction. This behavior of the velocity field causes
the bulk viscosity $\Pi $ to become piled up, which heats the matter in this
region. If this happens, the gradient of $\Pi $ becomes dominant in the
acceleration of the fluid compared to the pressure gradient (note that the
acceleration is given by $-\nabla \left( p+\Pi \right) $ ) since the
gradient of the pressure is proportional to $c_{s}$. Such a mechanism works
more efficiently in the direction where the initial acceleration is large,
and in consequence, in the direction to increase the elliptic flow.
Furthermore, this effect becomes effective only when significant collective
flow is formed near the phase transition. Thus, the recovery of $v_{2}$ by
this mechanism is expected for higher momentum particles. We consider that
this is the reason for the counter-intuitive behavior of our result. Of
course one should keep in mind that the applicability of the hydrodynamic
approach becomes dubious for the description of such a high $p_{T}$
dynamics so that such a mechanism may not necessarily be attributed directly
to experimentally observed values.

\begin{figure}[tbp]
\centering \includegraphics[scale=0.25]{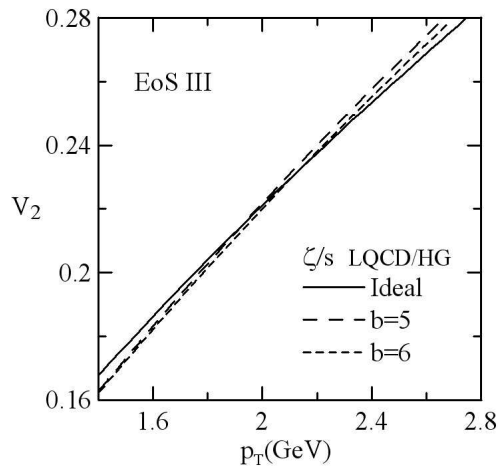}
\caption{$v_{2}$ as a function of $p_{T}$ for EoS III. The figure shows the
results for $\protect\zeta /s$ with critical behavior for different values
of $b_{\Pi }$, in the high transverse momentum region. The solid line corresponds
to the ideal result and the dotted and dashed lines correspond to $b=$ $6$
and $5$, respectively.}
\label{EoSII}
\end{figure}

So far, all the calculations presented were implemented fixing the
relaxation time parameter $b_{\Pi }=6$. This is because the $b_{\Pi }$
dependence for the EoS II is negligibly small. However, the $b_{\Pi }$
dependence can be observed for an EoS with a sharper phase transition such
as EoS III. In Fig. \ref{EoSII}, we show the elliptic flow calculated with
the EoS III for different relaxation times, $b_{\Pi }=5$ and $6$. One can
clearly see a dependence on the parameter $b_{\Pi }$ at high $p_{T}>1.6$ GeV although the curves of $v_{2}$ are independent of
the values of $b_{\Pi }$ for small $p_{T}<1.6GeV$.

\section{Concluding remarks}

\label{sec5}

In this work, we examined the effect of bulk viscosity on $v_{2}$ taking
into account the critical behavior of the EoS and transport coefficients
near the QCD phase transition. We found that the $p_{T}$ dependence of $v_{2}
$ is quantitatively changed by the presence of the QCD phase transition.
Within reasonable values of the transport coefficients, $v_{2}$ decreases by
a factor of $15\%$ at low $p_{T}$ ($<1$ GeV). However, for
larger values of $p_{T}$ ($>2$ GeV), the interplay between $%
c_{s}$ and $\zeta /s$ near the QCD\ phase transition generates more
collective flow, with even larger values than those of the ideal fluid.
These effects can depend on the value of the transport coefficient $b_{\Pi }$
when the phase transition is more abrupt. Of course one should keep in mind
that the applicability of the hydrodynamic approach is not guaranteed for
the high $p_{T}$ dynamics, but this is an interesting feature and deserves
further investigation.

All the calculations of $v_{2}$ here were done without dissipative
corrections to the one-particle distribution function. As discussed in this
paper, the naive application of the usual 14 moment calculation for the bulk
viscosity leads to unacceptable values for the corrections to the
distribution function. It should be noted that the 14 moments approximation
is not an unique choice to solve Grad's method and there are other
approaches to calculate the non-equilibrium distribution function \cite{guy}%
. It may be possible that the problem presented here can be solved using
alternative approaches. Further investigation on this aspect is required
urgently.

Although not explored in the present paper, one interesting aspect observed
in these studies should be mentioned. For much larger values of $\zeta /s$,
say $>0.5$, and smaller $b_{\Pi }$ ($<3$), we
found some peculiar behavior of $v_{2}$ at large $p_{T}$, exhibiting large
fluctuations for small changes in the transport coefficients. We identified
that such fluctuations are related to the emergence of instabilities in the
fluid evolution and that they are amplified when a sharp phase transition
(such as a first order phase transition) is present. We also found that
these instabilities cannot be controlled even after introducing the
additional viscosity discussed in \cite{dkkm2}. Although the shear viscosity
was not considered here, we believe that these instabilities are an
intrinsic phenomenon that appears for large bulk viscosity dynamics, such as
the Burgers instabilities \cite{dkkm3,mishust,burgers}. If this is true,
they should reflect in the observed $v_{2}$. For instance, the
event-by-event fluctuations in $v_{2}$ would become very large at high $p_{T}
$. Since this effect is closely related to the nature of the phase
transition, it would be interesting to verify this experimentally.

We acknowledge illuminating discussions from T. Hirano and A. Monnai and
thank J. Noronha for constructive comments
on the manuscript. This work has been supported by CNPq, FAPERJ, CAPES, PRONEX and the Helmholtz
International Center for FAIR within the framework of the LOEWE program
(Landesoffensive zur Entwicklung Wissenschaftlich- Okonomischer Exzellenz)
launched by the State of Hessen.

\appendix 

\section{14 Moments approximation}

\label{app:a}

In this appendix, we discuss the Grad's method with the 14 moments
approximation \cite{IS}. In this approximation, the non-equilibrium
one-particle distribution function is assumed as 
\begin{eqnarray}
f &=&\left( \exp \left( -y\right) +a\right) ^{-1}, \\
y &=&\left( \epsilon +\alpha _{0}\right) +K^{\mu }\left( \epsilon _{\mu
}-\beta _{0}u_{\mu }\right) +K^{\mu }K^{\nu }\epsilon _{\mu \nu },
\end{eqnarray}%
where the parameters $\epsilon $, $\epsilon _{\mu }$ and $\epsilon _{\mu \nu
}$ represent the deviation from equilibrium up to second order in momentum
and $a$ is $-1$ for bosons, $1$ for fermions and $0$ for a Boltzmann gas.
The index {0} is applied for the quantities in equilibrium. Without loss of
generality $\epsilon _{\mu \nu }$ is assumed to be traceless,%
\begin{equation}
\epsilon _{\mu }^{\mu }=0.
\end{equation}

For small momentum, the non-equilibrium one-particle distribution function
can be expanded around the equilibrium state 
\begin{equation}
f\approx f_{0}+f_{0}\left( 1-af_{0}\right) \left( y-y_{0}\right) +O\left(
2\right) ,  \label{distribution function}
\end{equation}%
where%
\begin{eqnarray}
f_{0} &=&\left( \exp \left( \alpha _{0}\left( x,t\right) -\beta _{0}\left(
x,t\right) K^{\mu }u_{\mu }\left( x,t\right) \right) +a\right) ^{-1}, \\
y-y_{0} &=&\epsilon +K^{\mu }\epsilon _{\mu }+K^{\mu }K^{\nu }\epsilon _{\mu
\nu }\ll 1.
\end{eqnarray}

For a rarefied gas, the conserved number current and the energy-momentum
tensor are expressed as 
\begin{align}
N^{\mu }& ={\displaystyle\int }\frac{d^{3}\vec{K}}{\left( 2\pi \right) ^{3}E}%
K^{\mu }f\left( K,T,\mu \right) ,  \label{Nmu_Boltz} \\
T^{\mu \nu }& ={\displaystyle\int }\frac{d^{3}\vec{K}}{\left( 2\pi \right)
^{3}E}K^{\mu }K^{\nu }f\left( K,T,\mu \right) .  \label{Tmunu_Boltz}
\end{align}%
Substituting Eq. (\ref{distribution function}) into them, we have 
\begin{align}
N^{\mu }& =I_{0}^{\mu }+\epsilon J_{0}^{\mu }+J_{0}^{\mu \nu }\epsilon _{\nu
}+J_{0}^{\mu \nu \lambda }\epsilon _{\nu \lambda }, \\
T^{\mu \nu }& =I_{0}^{\mu \nu }+\epsilon J_{0}^{\mu \nu }+J_{0}^{\mu \nu
\lambda }\epsilon _{\lambda }+J_{0}^{\mu \nu \lambda \rho }\epsilon
_{\lambda \rho },
\end{align}%
where,%
\begin{align}
I_{0}^{\alpha _{1}\ldots \alpha _{n}}& \equiv {\displaystyle\int }\frac{d^{3}%
\vec{K}}{\left( 2\pi \right) ^{3}E}K^{\alpha _{1}}\ldots K^{\alpha
_{n}}f_{0},  \notag \\
J_{0}^{\alpha _{1}\ldots \alpha _{n}}& \equiv {\displaystyle\int }\frac{d^{3}%
\vec{K}}{\left( 2\pi \right) ^{3}E}K^{\alpha _{1}}\ldots K^{\alpha
_{n}}f_{0}\left( 1-af_{0}\right) .  \label{moments}
\end{align}%
The 0-th order terms are identified as the equilibrium currents, 
\begin{align}
I_{0}^{\mu }& =N_{0}^{\mu }, \\
I_{0}^{\mu \nu }& =T_{0}^{\mu \nu }.
\end{align}%
All the remaining terms come from the dissipative corrections. The moments (%
\ref{moments}) can be expanded as 
\begin{align}
I_{0}^{\mu }& =I_{10}u^{\mu },  \notag \\
I_{0}^{\mu \nu }& =I_{20}u^{\mu }u^{\nu }-I_{21}\Delta ^{\mu \nu },  \notag
\\
J_{0}^{\mu }& =J_{10}u^{\mu },  \notag \\
J_{0}^{\mu \nu }& =J_{20}u^{\mu }u^{\nu }-J_{21}\Delta ^{\mu \nu },  \notag
\\
J_{0}^{\mu \nu \lambda }& =J_{30}u^{\mu }u^{\nu }u^{\lambda }-J_{31}u^{\mu
}\Delta ^{\nu \lambda }-J_{31}u^{\lambda }\Delta ^{\nu \mu }-J_{31}u^{\nu
}\Delta ^{\lambda \mu }  \notag \\
J_{0}^{\mu \nu \lambda \rho }& =J_{40}u^{\mu }u^{\nu }u^{\lambda }u^{\rho
}-J_{41}u^{\mu }u^{\nu }\Delta ^{\lambda \rho }-J_{41}u^{\mu }u^{\lambda
}\Delta ^{\nu \rho }-J_{41}u^{\mu }u^{\rho }\Delta ^{\lambda \nu
}-J_{41}u^{\nu }u^{\lambda }\Delta ^{\mu \rho }  \notag \\
& -J_{41}u^{\nu }u^{\rho }\Delta ^{\mu \lambda }-J_{41}u^{\lambda }u^{\rho
}\Delta ^{\mu \nu }+J_{42}\Delta ^{\mu \nu }\Delta ^{\lambda \rho
}+J_{42}\Delta ^{\mu \lambda }\Delta ^{\nu \rho }+J_{42}\Delta ^{\mu \rho
}\Delta ^{\nu \lambda },  \label{moments_exp2}
\end{align}%
where 
\begin{align}
I_{nq}& =\frac{1}{\left( 2q+1\right) !!}{\displaystyle\int }\frac{d^{3}\vec{K%
}}{\left( 2\pi \right) ^{3}E}E^{n-2q}\left( k^{2}\right) ^{q}f_{0}, \\
J_{nq}& =\frac{1}{\left( 2q+1\right) !!}{\displaystyle\int }\frac{d^{3}\vec{K%
}}{\left( 2\pi \right) ^{3}E}E^{n-2q}\left( k^{2}\right) ^{q}f_{0}\left(
1-af_{0}\right) .
\end{align}

In hydrodynamics, the dissipative corrections in the currents are
parameterized in terms of the bulk viscosity, the shear viscosity and the
heat conductivity. The variables $\epsilon $, $\epsilon _{\mu }$ and $%
\epsilon _{\mu \nu }$ can be determined in terms of these hydrodynamic
variables with the following set of constraints 
\begin{gather}
q_{\nu }=\Delta _{\mu \nu }N^{\mu },  \label{cond1} \\
\pi _{\mu \nu }=P_{\mu \nu \alpha \beta }\left( T^{\alpha \beta
}-T_{0}^{\alpha \beta }\right) ,  \label{cond2} \\
\Pi =-\frac{1}{3}\Delta _{\mu \nu }\left( T^{\mu \nu }-T_{0}^{\mu \nu
}\right) ,  \label{cond3} \\
u_{\mu }\left( N^{\mu }-N_{0}^{\mu }\right) =0,  \label{cond4} \\
u_{\nu }\left( T^{\mu \nu }-T_{0}^{\mu \nu }\right) =0.  \label{cond5}
\end{gather}

The first three relations define the irreversible currents while the last
two restrictions come from the Landau choice for the local equilibrium
reference frame. If we were using the Eckart frame, for example, we would
have to use, instead of (\ref{cond4}) and (\ref{cond5}), 
\begin{gather}
\left( N^{\mu }-N_{0}^{\mu }\right) =0,  \label{A1} \\
u_{\mu }u_{\nu }\left( T^{\mu \nu }-T_{0}^{\mu \nu }\right) =0.  \label{A2}
\end{gather}

The conditions (\ref{cond1}), (\ref{cond2}) and (\ref{cond3}) imply, 
\begin{align}
q^{\mu }& =-J_{21}\Delta ^{\mu \nu }\epsilon _{\nu }-2J_{31}\Delta ^{\mu \nu
}u^{\lambda }\epsilon _{\nu \lambda },  \label{from_cond_1} \\
\pi _{\mu \nu }& =2J_{42}P_{\mu \nu \lambda \rho }\epsilon ^{\lambda \rho },
\label{from_cond_2} \\
\Pi & =\epsilon J_{21}+J_{31}u^{\lambda }\epsilon _{\lambda
}+J_{41}u^{\lambda }u^{\rho }\epsilon _{\lambda \rho }-\frac{5}{3}%
J_{42}\Delta ^{\lambda \rho }\epsilon _{\lambda \rho }.  \label{from_cond_3}
\end{align}%
A general solution can be constructed using the following \textit{ansatz}, 
\begin{align}
\epsilon & =E_{0}\Pi , \\
\epsilon _{\lambda }& =D_{0}\Pi u_{\lambda }+D_{1}q_{\lambda }, \\
\epsilon _{\lambda \rho }& =B_{0}\left( \Delta _{\lambda \rho }-3u_{\lambda
}u_{\rho }\right) \Pi +B_{1}\left( u_{\lambda }q_{\rho }+u_{\rho }q_{\lambda
}\right) +B_{2}\pi _{\lambda \rho }.  \label{lastname}
\end{align}%
In order to know the form of the one-particle distribution function (\ref%
{distribution function}), we have to determine the form of the coefficients $%
E_{0}$, $D_{0}$, $D_{1}$, $B_{0}$, $B_{1}$ and $B_{2}$. This can be done by
substituting these general solutions into the equations discussed above,%
\begin{eqnarray}
\left( J_{21}+J_{31}\right) D_{1}+\left( J_{31}+J_{41}\right) B_{1} &=&-1,
\label{constr1} \\
B_{2} &=&\frac{1}{2J_{42}},  \label{constr2} \\
J_{21}E_{0}+J_{31}D_{0}+\left( 3J_{41}+5J_{42}\right) B_{0} &=&1,
\label{constr3} \\
J_{10}E_{0}+J_{20}D_{0}+\left( J_{30}+J_{31}\right) 3B_{0} &=&0,
\label{constr4} \\
J_{31}D_{1}+J_{41}B_{1} &=&0,  \label{constr5} \\
J_{20}E_{0}+J_{30}D_{0}+\left( J_{40}+J_{41}\right) 3B_{0} &=&0.
\label{constr6}
\end{eqnarray}

The first three equations, (\ref{constr1}), (\ref{constr2}) and (\ref%
{constr3}), come directly from (\ref{from_cond_1}), (\ref{from_cond_2}) and (%
\ref{from_cond_3}), respectively. The last three equations, (\ref{constr4}),
(\ref{constr5}), and (\ref{constr6}), are consequences of Eqs. (\ref{cond4})
and (\ref{cond5}). The solution of this set of equations is 
\begin{eqnarray}
B_{0} &=&\frac{1}{-3C_{1}J_{21}-3C_{2}J_{31}-3J_{41}-5J_{42}},
\label{final3} \\
B_{1} &=&\frac{J_{31}}{J_{41}J_{21}-J_{31}J_{31}}, \\
B_{2} &=&\frac{1}{2J_{41}}, \\
\frac{D_{0}}{3B_{0}} &=&-4\frac{J_{31}J_{20}-J_{41}J_{10}}{%
J_{30}J_{10}-J_{20}J_{20}}\equiv -C_{2},  \label{final1} \\
D_{1} &=&\frac{-J_{41}}{J_{41}J_{21}-J_{31}J_{31}}, \\
\frac{E_{0}}{3B_{0}} &=&m^{2}+4\frac{J_{31}J_{30}-J_{41}J_{20}}{%
J_{30}J_{10}-J_{20}J_{20}}\equiv -C_{1},  \label{final2}
\end{eqnarray}%
where $m$ is the mass of the particle.

In this paper, we are interested only in the bulk viscosity. Then, the form
of the distribution function is%
\begin{equation}
f=f_{0}+f_{0}(1-af_{0})(E_{0}+D_{0}(p^{\mu }u_{\mu })+B_{0}(p^{2}-4(p^{\mu
}u_{\mu })^{2}))\Pi ,
\end{equation}

We remark that the 14 moments approximation was formulated for a single
component gas. The generalization of this method for multiple components
systems has not yet been done.

\end{document}